\documentclass[aps,preprint,showpacs]{revtex4}
\usepackage{graphicx}
\usepackage{dcolumn}
\newcommand{\gmu}{{\gamma_\mu}}

\newcommand{\tauiso}{{\mbox{\boldmath $\tau$}}}
\newcommand{\bm}{\bibitem}

\begin{document}
                                                                                    
\title{$\eta$ meson production in nucleon-nucleon collisions within 
an effective Lagrangian model }
\author{R. Shyam}
\affiliation{Saha Institute of Nuclear Physics, Kolkata 70064, India }
                                                                                    
\date{\today}
                                                                                    
\begin{abstract}
                                                                                    
We investigate the $pp \to pp \eta$ and $pn \to pn \eta$ reactions
within an effective Lagrangian model for laboratory kinetic energies ranging 
from very close to the eta meson production threshold to about 10 GeV. 
Production amplitudes include contributions from the mechanism of excitation, 
propagation and decay of $N^*(1535)$, $N^*(1650)$, and $N^*(1710)$ baryonic 
resonances. The initial interaction between two incoming nucleons is modeled 
by the exchange of $\pi$, $\rho$, $\omega$ and $\sigma$ mesons where the 
vertex parameters are taken to be the same as those used in the
previous applications of this model. Parameters of the resonance vertices 
also have been taken from our earlier studies wherever applicable.  
Calculations have been done for total as well as differential $\eta$ 
production cross sections. 
To describe the data for energies closer to the production threshold 
final state interactions among the out-going particles have been included
by means of a generalized Watson-Migdal method. Terms corresponding to the 
excitation of $N^*$(1535) resonance and pion exchange process dominate the 
cross sections. With a single set of vertex parameters our model is 
able to describe the available data well on total cross sections for
beam energies ranging from close to threshold to upto 10 GeV. 
  
\end{abstract}
\maketitle

\newpage
\section{Introduction}
The low energy behavior of the quantum chromodynamics (QCD) is not
accessible to the perturbative approaches; the lattice gauge theory
\cite{wil74} is the ideal tool for this purpose. Despite enormous
computational power necessary for the numerical realization, lattice QCD
calculations have started, very recently, to describe masses and other 
constants of the baryonic ground as well as excited states~\cite{lei05}.
Experimentally, the determination of baryonic resonance properties proceeds
indirectly by exciting the nucleon with the help of a hadronic or 
electromagnetic probe and performing measurements for their decay products
(mesons and nucleons). The reliable extraction of nucleon resonance properties
from such experiments is a major issue of the hadron physics.

In recent years, important advances have been made in the experimental 
investigation of meson production reactions in nucleon-nucleon ($NN$)
collisions, particularly at beam energies very close to respective production
thresholds
~\cite{mos02,mey01,bal98,sew99,kow04,bar04,sam06,cal98,cal99,smy00,bar03,bal04,mos04}.
Low incident energies also provide the opportunity to investigate
the meson-nucleon interactions through these reactions since in this energy
regime the final state interactions among the outgoing particles affect 
strongly the meson production cross sections.

The $\eta$ meson, which is the next lightest nonstrange member in the meson
mass spectrum, has been a subject of considerable interest.  It has 
been thought of as a probe for the $s{\bar s}$ component in the nucleon
wave function~\cite{dov90}. There is also interest in measuring the
rare decays of $\eta$ which could provide a new rigorous test of the
standard model~\cite{pap05} or even of the physics beyond this. The nucleon
resonance $N^*$(1535) [$S_{11}(1535)$] with spin $1\over 2$, isospin 
$1\over 2$, and odd parity, has a remarkably large $\eta N$ branching ratio.
It lies very close to the threshold of the $NN \to NN\eta$ reaction and 
contributes to the amplitude of this reaction even at the threshold. 
Therefore, the study of $\eta$-meson production in $NN$ collisions at the 
near threshold beam energies provides the unique opportunity to 
investigate the properties of $N^*$(1535) which have been the subject 
of some debate recently (see, e.g.,~\cite{mat05}). The attractive nature
of the $\eta$-nucleon interaction may lead to the formation of bound 
(quasi-bound) $\eta$-nucleus states (see, e.g.,
\cite{chi91,fix02,hai02,pfe04,sib04}). This subject has been a topic of
intense discussion at a recent work shop~\cite{eta06}.

Production of $\eta$ meson in heavy ion collisions also is of great interest.
Due to the high threshold of the elementary production reaction, 
$\eta$ mesons in such collisions are produced only by very energetic nucleons
and reflect therefore, the tails of the nucleon momentum distributions as 
they arise in a high density and high temperature phase of the collision
\cite{cas90}. The elementary $NN\eta$ production cross sections are a 
crucial ingredient in the transport model studies of the $\eta$-meson
production in the nucleus-nucleus collisions.

Since lattice QCD calculations are still far from being amenable to the
low and intermediate energy scattering and reaction processes,
the effective methods are mostly used for the description of the 
dynamics of the meson production reactions in hadronic collisions. These 
approaches introduce the baryonic resonance states explicitly in their 
framework and try to extract their properties by comparing
the theoretical results with experimental observables. Several authors have
used models of such type in describing the $\eta$ meson production in $NN$ 
collisions~\cite{vet91,moa96,fal01,nak02,nak03,var04,fix04}. 

The main objective of this paper is to investigate the $\eta$ meson 
production in $NN$ collisions in the framework of an 
effective Lagrangian Model (ELM) which has been used earlier rather 
successfully to describe the pion~\cite{eng96,shy98}, associate 
kaon~\cite{shy99}, and dilepton~\cite{shy03} production data in such 
collisions. The motivation here is to see as to how far can one explain 
the recently measured data on total and differential cross sections  
of $pp\to pp \eta$~\cite{cal99,smy00,bar03,bal04,mos04} and  
the $pn \to pn \eta$~\cite{cal98} reactions within this model using
same sets of entrance channel and resonance channel parameters
(for those resonances that appeared in earlier applications
\cite{shy99,shy03}).

Within the ELM, initial interaction between two incoming nucleons is 
modeled by an effective Lagrangian which is based on the exchange
of $\pi$, $\rho$, $\omega$ and $\sigma$ mesons. The coupling
constants at the nucleon-nucleon-meson vertices are determined by 
directly fitting the $T$ matrices of the $NN$ scattering
in the relevant energy region~\cite{sch94}. The effective Lagrangian
uses the pseudovector (PV) coupling for the nucleon-nucleon-pion
vertex, as it is consistent with the chiral symmetry requirement of 
the QCD~\cite{wei68,ber95,han04} and also it leads to negligible 
contributions from the negative energy states ("pair suppression 
phenomena")~\cite{mac87}. The $\eta$ meson  production proceeds via 
excitation of $N^*$(1535), $N^*$(1650) and $N^*$(1710) intermediate
baryonic resonance states which have known branching ratios for the 
decay into the $\eta N$ channel. The coupling constants at the 
resonance-nucleon meson vertices are determined from the experimental 
widths for the decay of the resonances into the relevant channels 
except for those involving the $\omega$ meson where they are determined 
from the vector meson dominance (VMD) hypothesis. The interference terms
between amplitudes corresponding to various meson exchanges and the 
intermediate resonance states have been included.

The final state interaction (FSI) among the outgoing particles affect 
strongly the cross sections of the $NN \to NN\eta$ reaction at near 
threshold beam energies~\cite{nak02,nak03,fix04}. In applications of the ELM
to describe the near threshold meson production reactions in $NN$
collisions~\cite{shy98,shy99}, the FSI effects were included within the
Watson-Midgal theory~\cite{wat69} which is based on the assumption that
the FSI effects are strong in relation to the production process and that
they occur attractively between only one particular pair of out-going particles.
In Ref.~\cite{shy99}, this method was used somehow arbitrarily for all
the 3 outgoing pairs of particles. However, Watson's method as such is not
applicable, in strict sense, for those cases where the attraction between
outgoing particles is not so pronounced or where interaction between more
than one pair is to be included in calculations. In this paper, we 
employ a generalized Watson method in which three-body states are treated
by splitting the total interaction into pairwise net interactions which
leads to a series decomposition of the net scattering among all the
particles in terms of separate total scattering between  pairs of particles
(see, e.g.~\cite{gil64}). However, the three-body interactions are 
neglected. In view of the arguments presented in Refs.~\cite{delp04,fix04}
in favor of using the three-body scattering theory to describe the 
$NN\eta$ process, it would be interesting to see to what extent this
generalized method is able to explain the $NN\eta$ production data.

In the next section, we present a brief description of the ELM where we 
describe the main ingredients of the theory and give all the input parameters
used in our calculations. The generalized Watson method of FSI effects is
also described in a subsection here. The results of calculations and
their discussions are presented in section III. Summary and conclusions
of our work are given in section IV. 

\section{Formalism}

A representative of the lowest order Feynman diagrams contributing to
the $\eta$ meson production considered by us, is shown in Fig.~1.
Momenta of various particles are indicated in Fig.~1a. $q$, $p_i$, and
$p_\eta$ are four momenta of the exchanged meson, the intermediate 
resonance and the $\eta$-meson, respectively. To evaluate various 
amplitudes, we have used the effective Lagrangians for the 
nucleon-nucleon-meson and resonance-nucleon-meson vertices as described
below.  
\begin{figure}
\begin{center}
\includegraphics[width=0.4 \textwidth]{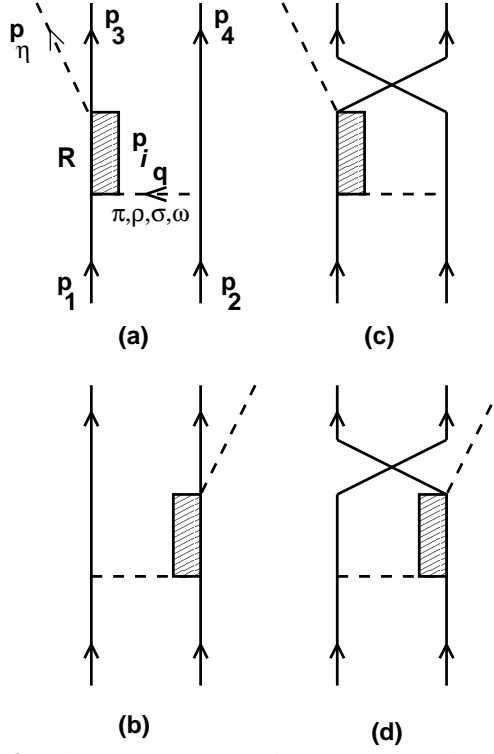}
\vskip -0.3in
\caption{
Feynman diagrams for the $\eta$-meson production in nucleon-nucleon
collisions. Diagrams (a) and (b) show the direct processes, while (c)
and (d) the exchange ones. $R$ represents a baryonic resonance.
}
\end{center}
\end{figure}

\subsection{Nucleon-nucleon-meson vertex}
                                                                                
As done before in the investigation of $pp \rightarrow pp\pi^0$,
$pp \rightarrow pn\pi^+$~\cite{eng96}, $pp \rightarrow p \Lambda 
K^+$~\cite{shy99}, and $NN \to NN e^+e^-$~\cite{shy03} reactions, 
the parameters for nucleon-nucleon-meson vertices are determined by
fitting the $NN$ elastic scattering $T$ matrix with an effective $NN$
interaction based on the $\pi$, $\rho$, $\omega$, and $\sigma$ meson 
exchanges. The effective $NNmeson$ Lagrangians are 
(see, e.g.,~\cite{wis88,pen02})
\begin{eqnarray}
{\cal L}_{NN\pi} & = & -\frac{g_{NN\pi}}{2m_N} {\bar{\Psi}}_N \gamma _5
                             {\gamma}_{\mu} \tauiso
                            \cdot (\partial ^\mu {\bf \Phi}_\pi) \Psi _N. \\
{\cal L}_{NN\rho} &=&- g_{NN\rho} \bar{\Psi}_N \left( \gmu + \frac{k_\rho}
                         {2 m_N} \sigma_{\mu\nu} \partial^\nu\right)
                          \tauiso \cdot \mbox{\boldmath $\rho$}^\mu \Psi_N. \\
{\cal L}_{NN\omega} &=&- g_{NN\omega} \bar{\Psi}_N \left( \gmu + \frac{k_\omega}                         {2 m_N} \sigma_{\mu\nu} \partial^\nu\right)
                          \omega^\mu \Psi_N.   \\
{\cal L}_{NN\sigma} &=& g_{NN\sigma} \bar{\Psi}_N \sigma \Psi_N.
\end{eqnarray}
In Eqs.~(1)-(4), we have used the notations and conventions of
Bjorken and Drell~\cite{bjo64} and definitions of various terms
are the same as those given there. In Eq.~(1) $m_N$ denotes
the nucleon mass.
It should be noted that we use a PV coupling for the $NN\pi$ vertex.
Since these Lagrangians are used to directly model the $NN$ $T$ matrix, we
have also included a nucleon-nucleon-axial-vector-isovector vertex,
with the effective Lagrangian given by
\begin{eqnarray}
{\cal L}_{NNA} & = & g_{NNA} {\bar {\Psi}} \gamma_5 \gamma_\mu \tauiso \Psi
                     \cdot {\bf {A}}^\mu,
\end{eqnarray}
where ${\bf A}$ represents the axial-vector meson field.
This term is introduced because if the mass of the axial meson $A$ is 
chosen to be very large ($\gg m_N$)~\cite{sch94} and $g_{NNA}$ is defined as
\begin{eqnarray}
g_{NNA} & = & \frac{1}{\sqrt{3}} m_A \left(\frac{g_{NN\pi}}{2m_N}\right),
\end{eqnarray}
it cures the unphysical behavior in the angular distribution of
$NN$ scattering caused by the contact term in the one-pion exchange
amplitude. It should be mentioned here that $A$ is different from
the $a_1$(1260) meson resonance. The role of the
$A$ vertex is to explicitly subtract out the contact term of the one-pion
exchange part of the $NN$ interaction. Similar term in the
coordinate space potential is effectively switched off by the repulsive
hard core.
                                                                                
At each interaction vertex, the following form factor is introduced
\begin{eqnarray}
F_{i}^{NN} & = & \left (\frac{\lambda_i^{2} - m_i^{2}}{\lambda_i^{2} - q_i^{2}}
        \right ), i= \pi, \rho, \sigma, \omega,
\end{eqnarray}
where $q_i$ and $m_i$ are the four momentum and mass of the $i$th
exchanged meson and $\lambda_i$ is the corresponding cut-off parameter.
The latter governs the range of suppression of the contributions of
high momenta which is done via the form factor. Since $NN$ scattering
cross sections decrease gradually with the beam energy (beyond certain value),
and since we fit the elastic $T$ matrix directly, the coupling
constants are expected to be energy dependent. Therefore, while keeping the
cut-offs $\lambda_i$ [in Eq.~(7)] energy independent,
we take energy dependent meson-nucleon coupling constants of the following
form
\begin{eqnarray}
g(\sqrt{s}) & = & g_{0} exp(-\ell \sqrt{s}),
\end{eqnarray}
where $s$ is the square of the total CM energy. The parameters $g_0$, $\ell$, 
and $\lambda$ were determined by fitting to the $T$ matrix of the relevant
proton-proton and proton-neutron scattering data at the
beam energies in the range of 800 MeV to 4.00 GeV~\cite{sch94}. This
procedure also fixes signs of the effective Lagrangians [Eqs. (1)-(5)].
The values of various parameters are shown in Table I [the signs of all the
coupling constants (g) are positive]. In this table the parameters of the
$A$ exchange vertex are not explicitly shown as they are related to those
of the pion via Eq.~(6). We would like to remark that the same parameters were 
also used to describe the initial $NN$ interaction in the calculations reported
in Refs.~\cite{eng96,shy99,shy03}. This ensures that the elastic $NN$ elastic 
scattering channel remains the
\begin{table}[here]
\begin{center}
\caption[T1]{ Coupling constants for the $NN$-meson vertices used in the
calculations}
\vspace{.5cm}
\begin{tabular}{|c|c|c|c|c|}
\hline
 Meson & $g^2/4\pi$ & $\ell$ & $\Lambda$ & mass \\
       &             &        & (\footnotesize{GeV} ) & (\footnotesize{GeV})
 \\ \hline
$\pi    $ & 12.562 & 0.1133 & 1.005 & 0.138 \\
$\sigma $ & 2.340  & 0.1070 & 1.952 & 0.550 \\
$\omega $ & 46.035 & 0.0985 & 0.984 & 0.783 \\
$\rho   $ & 0.317  & 0.1800 & 1.607 & 0.770 \\
$k_{\rho}$ = 6.033, $k_{\omega}$ = 0.0 & & & & \\ \hline
\end{tabular}
\end{center}
\end{table}
\noindent
same in the description of various inelastic processes
within this model. 
 
The  main criterion for choosing the meson exchanges as discussed above 
is to describe the $NN$ scattering in the relevant beam energy region. 
We have left out the $\eta$-meson exchange in our description of the $NN$
interaction since due to the psuedoscalar (PS) nature of its coupling, the
contribution of the exchange terms of this particle (having a mass much 
larger than that of the psuedoscalar pion) is expected to be very small.
Furthermore, as confirmed by several studies (see, 
e.g.,~\cite{gre80,ben95,cbn95,kir96}), the coupling constant for the 
$NN\eta$ vertex is rather small. 
 
\subsection{Resonance-nucleon-meson vertex}

As the $\eta$ meson has zero isospin, only isospin-${1 \over 2}$
nucleon resonances are allowed. Below 2 GeV center of mass (c.m.)
energy, $N^*$(1535) has a prominent decay branching ratio of 40-60$\%$
into the $N\eta$ channel (see, e.g., the latest review of the particle data
group~\cite{pdg06}). On the other hand, $N^*$(1650) and $N^*$(1710)
resonances have a small but non-negligible decay branching ratios of
3 - 10$\%$ and 6 $\pm$ 1$\%$, respectively, to this channel. In comparison
to these, the branching ratio for the decay of $N^*$(1520) resonance to 
$N\eta$ channel is negligibly small and we have not included it in out
description. In several previous studies of the $NN \to NN\eta$ reaction, 
contributions from only $N^*$(1535) resonance have been considered. 

Since all the three resonances can couple to the meson-nucleon channels 
considered in the previous section, we require the effective Lagrangians
for all the four resonance-nucleon-meson vertices corresponding to all
the included resonances. At the spin-${1 \over 2}$ resonances - $N-\pi$
($\eta$) vertices, we have the choice of PS or PV couplings. The
corresponding effective Lagrangians can be written 
as~\cite{ben95,feu97,feu98,shy99}
\begin{eqnarray}
{\cal L}^{PV}_{N_{1/2}^*N\pi} & = & -\frac{g_{N_{1/2}^*N\pi}}{M}
                          {\bar{\Psi}}_{N^*} {\Gamma}_{\mu} \tauiso
                       \cdot (\partial ^\mu {\bf \Phi}_\pi) \Psi _N
                          + {\rm h.c.}\\ 
{\cal L}^{PS}_{N_{1/2}^*N\pi} & = & -g_{N_{1/2}^*N\pi}
                          {\bar{\Psi}}_{N^*} i{\Gamma} \tauiso
                        {\bf \Phi}_\pi \Psi _N + {\rm h.c.},\\ 
{\cal L}^{PV}_{N_{1/2}^*N\eta} & = &
                    -\frac{g_{N^*_{1/2}\eta}}{M}
                          {\bar{\Psi}}_{N^*} {\Gamma}_{\mu} \tauiso
                       \cdot (\partial ^\mu {\bf \Phi}_{\eta}) \Psi _N
                        + {\rm h.c.},\\ 
{\cal L}^{PS}_{N_{1/2}^*N\eta} & = & -g_{N^*_{1/2}N\eta}
                          {\bar{\Psi}}_{N^*} {i\Gamma} \tauiso
                        {\bf \Phi}_{\eta} \Psi _N + {\rm h.c.}, 
\end{eqnarray}
where $M \,=\,(m_{N^*}\,\pm\,m_N)$, with upper sign for even parity and
lower sign for odd parity resonance. The operators $\Gamma$, $\Gamma_\mu$,
are given by,
\begin{eqnarray}
\Gamma = \gamma_5,\,\, \Gamma_\mu = \gamma_5 \gamma_\mu,\\
\Gamma = 1, \,\, \Gamma_\mu = \gamma_\mu,
\end{eqnarray}
for resonances of even and odd parities, respectively. 
We have performed calculations with both of these couplings. The effective
Lagrangians for the coupling of resonances to other mesons are,
\begin{eqnarray}
{\cal L}_{N_{1/2}^*N\rho} & = &- g_{N_{1/2}^*N\rho} \bar{\Psi}_{N^*}
                           \frac{1} {2m_N}
                            \Gamma_{\mu\nu} \partial^\nu
                      \tauiso \cdot \mbox{\boldmath $\rho$}^\mu \Psi_N. +
                         {\rm h.c.}\\
{\cal L}_{N_{1/2}^*N\omega} &=&- g_{N_{1/2}^*N\omega} \bar{\Psi}_{N^*}
                         \frac{1}{2m_N}
                         \Gamma_{\mu\nu} \partial^\nu
                          \omega^\mu \Psi_N. + {\rm h.c.}  \\
{\cal L}_{N_{1/2}^*N\sigma} & = & g_{N_{1/2}^*N\sigma} \bar{\Psi}_{N^*}
                             \Gamma^\prime\sigma \Psi_N + {\rm h.c.},
\end{eqnarray}
where operators ${\Gamma^\prime}$ and $\Gamma_{\mu\nu}$ are,
\begin{eqnarray}
\Gamma^\prime = 1, \,\, \Gamma_{\mu\nu} = \sigma_{\mu\nu}\\
\Gamma^\prime = \gamma_5,\, \, \Gamma_{\mu\nu} = \gamma_5 \sigma_{\mu\nu},
\end{eqnarray}
for resonances of even and odd parities, respectively.

We assume that the off-shell dependence of the $NN^*$ vertices are determined
solely by multiplying the vertex constants by  form factors. Similar to 
Refs.~\cite{pen02,shy06}, we use the following form factors for $N^*N$-meson 
vertices  
\begin{eqnarray}
F_{i}^{NN^*} & = & \left [\frac{(\lambda_i^{N^*})^4}
                   {(\lambda_i^{N^*})^4 + ( q_i^{2} -m_i^2)^2}
                    \right ], i= \pi, \rho, \sigma, \omega,
\end{eqnarray}

The resonance couplings are determined from the experimentally 
observed branching ratios for the decay of the
resonances to the corresponding channels. Since the resonances considered
in this study have no known branching ratios for the decay into the 
$N\omega$ channel, we determine the coupling constants for the 
$N^*N\omega$ vertices by the strict vector meson dominance (VMD) 
hypothesis~\cite{sak69}, which  is based essentially on the assumption 
that the coupling of photons to hadrons takes place through a vector 
meson. For details of these calculations we refer to~\cite{shy99}.

The resonance properties and the values of various coupling constants
are given in Table II. Value of the cut-off parameter ($\lambda_i^{N^*}$)
is taken to be 1.2 GeV for all the vertices, which is the same as that 
used in Refs.~\cite{shy03,shy06}. Fixing $\lambda_i^{N^*}$ to one value
minimizes the number of free parameters.
\begin{table}[here]
\begin{center}
\caption {Resonance parameters and the coupling constants for various decay
vertices. Coupling constants at the $N^*N\omega$ vertices are obtained from
the vector meson dominance hypothesis (see, {\it e.g.} Ref.~\cite{shy99}).}
\vspace{0.5cm}
\begin{tabular}{|c|c|c|c|}
\hline
Resonance  & Width & Decay channel  & $g$ \\
           & (GeV) &                &    \\
\hline
$N^*$(1535)& 0.150 & $N\pi$         & 0.6840  \\
           &       & $N\rho$        & 3.9497  \\
           &       & $N\omega$      & 1.4542  \\
           &       & $N\sigma$      & 2.5032  \\
           &       & $N\eta  $      & 2.2000  \\
$N^*$(1650)& 0.150 & $N\pi$         & 0.8096  \\
           &       & $N\rho$        & 2.6163  \\
           &       & $N\omega$      & 1.8013  \\
           &       & $N\sigma$      & 2.5032  \\
           &       & $N\eta$        & -0.5469 \\
$N^*$(1710)& 0.150 & $N\pi$         & 1.0414  \\
           &       & $N\rho$        & 2.9343  \\
           &       & $N\omega$      & 1.5613  \\
           &       & $N\sigma$      & 0.6737  \\
           &       & $N\eta$        & 1.0328  \\
\hline
\end{tabular}
\end{center}
\end{table}

It should however, be stresses that the branching ratios determine 
only the square of the corresponding coupling constants, thus their
signs remain uncertain in this method. Predictions from independent
studies are used as a guide to fix these signs. We have followed here
the results of Ref.~\cite{pen02} for this purpose. 
The propagators for various mesons and nucleon resonances in the
calculation of the amplitudes have been taken to be the same as those
discussed in~\cite{shy99,shy03}. 

\subsection{Amplitudes and cross sections}

After having established the effective Lagrangians, coupling constants
and form of the propagators, the amplitudes for various diagrams associated
with the $NN \to NN\eta$ reaction can be calculated in straight forward manner
by following the well known Feynman rules. The isospin part is treated 
separately. This gives rise to a constant factor for each graph, which
is shown in Table III.
\begin{table}
\begin{center}
\caption{Isospin factors for various diagrams, Isovector corresponds to
$\pi$ and $\rho$ exchange graphs while isoscalar to $\omega$ and $\sigma$ ones}
\vskip .1in
\begin{ruledtabular}
\begin{tabular}{|ccc|}
graph & isovector & isoscalar\\
\hline
      & {\bf $pp \to pp \eta$} &   \\
direct   & 1.0 & 1.0 \\
exchange & 1.0 & 1.0 \\
                                                                                
      & {\bf $pn \to pn\eta$} &   \\
                                                                                
direct   & -1 & 1 \\
exchange &  2 & 0 \\
\end{tabular}
\end{ruledtabular}
\end{center}
\end{table}

It should be stressed here that signs of various amplitudes are 
fixed, by those of the effective Lagrangians, coupling constants
and propagators as described above. These signs are not 
allowed to change anywhere in the calculations. 

The general formula for the invariant cross section of the $NN
\rightarrow NN\eta$ reaction is written as 
\begin{eqnarray}
d\sigma & = & \frac{m_N^4}{2\sqrt{[(p_1 \cdot p_2)^2-m_N^4]}}
                     \frac{1}{(2\pi)^5}\delta^4(P_f-P_i)|T_{fi}|^2
                     \prod_{a=1}^3 \frac{d^3p_a}{E_a},
\end{eqnarray}
where $T_{fi}$ represents the total amplitude, $P_i$ and $P_f$ the sum of
all the momenta in the initial and final states, respectively, and
$p_a$ the momenta of the three particles in the final state. The
corresponding cross sections in the laboratory or center of mass systems
can be written from this equation by imposing the relevant conditions.

\subsection{Final state interaction}

For describing the data for the $NN \rightarrow NN \eta$ reaction at beam 
energies very close to the $\eta$ production threshold, consideration of 
the final state interaction (FSI) among the three out going particles is
important. We follow here an approximate scheme in line with the 
Watson-Migdal theory of FSI~\cite{wat69}. In this approach the energy
dependence of the cross section due to FSI is separated from that of the
primary production amplitude. This is based on the assumption that the 
reaction takes place over a small region of space, a condition fulfilled
rather well in near threshold reactions involving heavy mesons. This method
has been extensively applied to study the low momentum behavior of the 
pion~\cite{dub86,shy98,mei98}, $\eta$ meson~\cite{mol96,dru97,del04}, 
associated hyperon~\cite{shy99,sib06} and $\phi$-meson~\cite{tit00} 
production in $NN$ collisions. The total amplitude is written as
\begin{eqnarray}
T_{fi} & = & T_0(NN \rightarrow NN\eta) \cdot T_{ff},
\end{eqnarray}
where $T_{0}(NN \rightarrow NN\eta)$ is the primary production
amplitude, while $T_{ff}$ describes the re-scattering 
among the final particles which goes to unity in the limit of no FSI. 
The factorization of the total amplitude into those of the FSI and primary
production (Eq.~(22)), enables one to pursue the diagrammatic approach
for the latter within an effective Lagrangian model and investigate
the role of various meson exchanges and resonances in describing
the reaction. 

Watson's original method~\cite{wat69} was developed for those cases where
the final state interaction is strong in relation to the production process
and where it is confined only to one particular pair of particles (mostly
among nucleons in case of nucleon-nucleon-meson final states). On the
other hand, in certain cases it may be necessary to include FSI among all 
the three outgoing particles since even if the meson-baryon interactions 
are weak, they can still be influential through interference. In 
Ref.~\cite{shy99}, the $T$ matrix $T_{ff}$ was written (without providing
any proof) as a coherent sum of the transition matrices describing the 
final state interaction among all three two-body subsystems of the final state 
nucleon-nucleon-meson system. We show here that this result can be obtained
(in a slightly different form) by following the technique of multiple FSI
as discussed in Ref.~\cite{gil64}. To that end, we first try to get a 
representation for the total amplitude $T_{fi}$ in terms of an expression 
similar to that given by Eq.~(22) where the FSI amplitude $T_{ff}$ is 
appropriately constructed. 

To introduce the treatment of the FSI for the three-particle system, we
assume that the three-particle final states can be represented by additive 
potentials of the form $U  =  U_{12} + U_{31} + U_{23} \equiv U_3 + U_2 + U_1$. 
With this assumption, the total amplitude $T_{fi}$ can be written as a 
iteration series in terms of the production amplitude $T_0$ (defined in 
Eq.~(22)) and the three-body final state pair interaction amplitudes 
$T_{ij} \equiv T_k$ (see Appendix A for details). 
\begin{eqnarray}
T_{fi} & = & T_0 + \sum_k T_k G_0 T_0 + \sum_{k \not= j} 
T_{k} G_0 T_{j} G_0 T_0 + ....,
\end{eqnarray}   
where the final state interaction transition matrices $T_k$ are as 
defined in Appendix A.  $G_0$ is the Green's function corresponding to 
the free Hamiltonian (kinetic energy). Neglecting processes depicted by
the third term, this decomposition can be expressed by Fig. 2.
\begin{figure}
\begin{center}
\includegraphics[width=0.5 \textwidth]{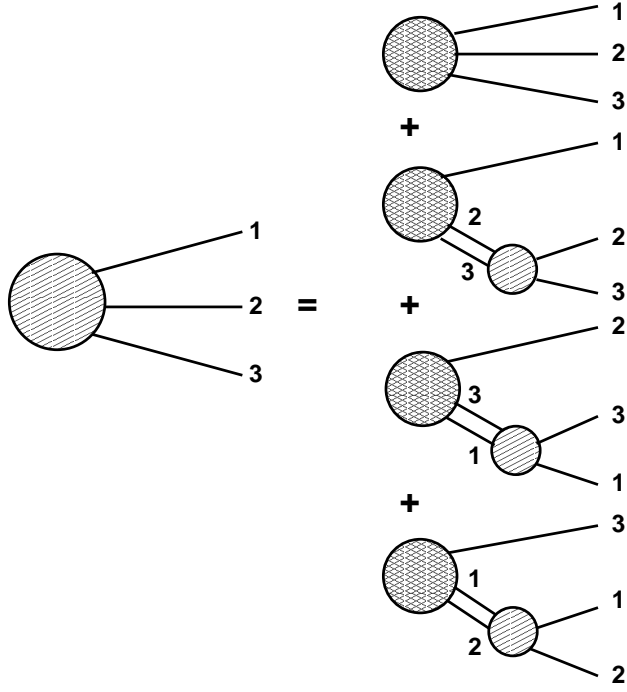}
\vskip -0.1in
\caption{
Final state scattering among three particles evaluated to the lowest order
}
\end{center}
\end{figure} 
It is easy to show that the result of the Watson FSI theory are recovered
if one retains only the amplitude $T_0$ and a single pair amplitude, say,
$T_{12}$.

Any practical calculation requires evaluation of the matrix element 
$<123|T_{12}G_0T_0|N_1N_2>$, where $N_1$ and $N_2$ denote two particles of
the incident channel and 1,2 and 3 represent the outgoing channel particles.
Introducing a complete set of intermediate states of particles, say, 
$1$ and $2$, we get for this channel
\begin{eqnarray}
T_{fi}  =  \frac{1}{(2\pi)^3} \int \frac{d^3{\bf k}_1^\prime}{2E_1^\prime}
\frac{d^3{\bf k}_2^\prime}{2E_2^\prime}\delta^3({\bf k}_1^\prime +
{\bf k}_2^\prime - {\bf k}_1 - {\bf k}_2) 
\frac{<{\bf k}_1{\bf k}_2|T_{12}|{\bf k}_1^\prime{\bf k}_2^\prime>
<{\bf k}_1^\prime{\bf k}_2^\prime {\bf k_3}|T_0|N_1N_2>}
{E-(E_3+E_1^\prime + E_2^\prime)+i\epsilon}.\nonumber\\
\end{eqnarray}
It may be noted that in this form, each particle lies on the mass shell
and three momentum is conserved in the intermediate processes. Introducing
the total and relative momenta, ${\bf p}^\prime = 
{\bf k}_1^\prime + {\bf k}_2^\prime$, $2q^\prime = 
{\bf k}_1^\prime - {\bf k}_2^\prime$, and evaluating the integral 
in the barycentric frame of 1 and 2, we obtain,
\begin{eqnarray}
T_{fi} & = & \frac{1}{(2\pi)^3} \int \frac{T_{12}(\xi,\theta;\xi^\prime,
           \theta^\prime)T_0(\xi^\prime,\theta^\prime,{\bf k}_3; {\bf k}_i)}
           {\xi - \xi^\prime + i\epsilon} \frac{2q^\prime d\xi^\prime
           d\Omega^\prime}{\xi^\prime},
\end{eqnarray}
where $\xi$ is the energy of $1$ and $2$ in this frame, $\xi^\prime$ is
the intermediate state and $\theta$ denotes the orientation $(\theta,\phi)$ of
{\bf q} with respect to a fixed axis.

For further evaluation of the integral, we make a partial wave 
decomposition of the amplitude and for each partial wave we rewrite it 
as 
\begin{eqnarray}  
T_{fi}^{\ell m} & \propto & T_{12}(\xi,\xi)T_0(\xi) \int 
\frac{[T_{12}(\xi,\xi^\prime)T_0(\xi^\prime)/T_{12}(\xi,\xi) T_0(\xi)]}
           {\xi - \xi^\prime + i\epsilon} \frac{2q^\prime d\xi^\prime
           d\Omega^\prime}{\xi^\prime},
\end{eqnarray}
Now we make the assumption that the ratio within the square brackets in 
the integrand of Eq.~(26) is constant upto a certain energy $\xi_c$ and
zero thereafter. Extending this procedure to all the three interacting 
pairs, we get in the low energy and $s$-wave limit  
\begin{eqnarray}
T_{fi} & \simeq & T_0(\xi) T_{ff}(\xi).
\end{eqnarray}
In Eq.~(27) $T_{ff}$ is defined as  
\begin{eqnarray}
T_{ff}(\xi) & = &\sum_{i\not=j}c_{ij}T_{ij}(\xi,\xi),
\end{eqnarray}
where
\begin{eqnarray}
c_{ij} & = & \frac{1}{\pi} cosh^{-1}
\left [ 1 + \frac{16m_im_j(\xi_c-m_i - m_j)}{(m_i + m_j)^3} \right ].
\end{eqnarray}
It is obvious that Eq.~(28) allows interference among the final state 
scattering amplitudes.  We further note that apart from the factor 
$c_{ij}$, this equation is similar to that used in Ref.~\cite{shy99}.  

The derivation of Eq.~(27) is independent of the strength of the interaction
and of whether it is attractive or repulsive. The quantity $c_{ij}$ can be
regarded as the amount of final state scattering that takes place in a
particular channel $ij$. A plausible value of the cut off $\xi_c$ comes
from the constraint that for the $NN$ sub-state $c_{ij}$ should come out to
be unity since in the limit of FSI in only this sub-state, we should
recover Watson's result. With the same value of the $\xi_c$,  the
$c_{ij}$ for the $\eta N$ sub-state comes out to be 1.07. It must
however, be noted that this procedure determines the value of the
$c_{ij}$ at best only for the $NN$ channel. It remains
largely undetermined for the $\eta - p$ sub-state which could even be
dependent on the relative energy of this channel. Since, for the time
being we do not have a definite method to determine this quantity, we
have taken the same value for the parameter $\xi_c$ for both the sub-states.

For calculating the FSI amplitude for the $\eta N$ sub-state, we note 
that there are no direct measurements of the elastic $\eta N$ scattering
and the information about the $\eta N$ elastic scattering amplitude is 
obtained by describing the $\pi N \to \eta N$ and $\gamma N \to \eta N$ 
data within some model. Recently, there has been suggestions to determine
the $\eta N$ scattering amplitude from the studies of associated 
photoproduction of $\phi$ and $\eta$ mesons off the proton~\cite{soy06}.
We adopt here the results reported in Ref.~\cite{gre05} where $\eta N$ 
scattering parameters have been obtained by fitting the $\pi N \to \pi N$,
$\pi N \to \eta N$, $\gamma N \to \eta N$ data in an energy range from 
threshold to about 100 MeV, in a $K$ matrix method. These authors write 
the elastic $\eta N$ scattering $T$ matrix as
\begin{eqnarray}
T^{-1} & = & 1/a + \frac{r_0}{2} q_\eta^2 + sq_\eta^4 - iq_\eta,
\end{eqnarray}
where $q_\eta$ is the momentum in the $\eta N$ center of mass (c.m.) system.
Seven sets of values of the parameters $a$, $r_0$ and $s$ are given in 
Ref.~\cite{gre05}. We found that the best description of the data (within
the realm of our overall input parameter sets given in Tables I and II), is  
provided by the $\eta N$ scattering amplitudes with the parameter set, 
$a = 0.51 + i 0.26$ fm, $r_0 = -2.50 - i 0.310$ fm, and 
$s = -0.20 - i 0.04$ fm$^3$.
It should be noted that the real part of the $\eta N$ scattering length
of this parameter set ($a_R$) is about half of that of the "preferred" 
set of Ref.~\cite{gre05}. A larger $a_R$ is also supported by the 
calculations presented in Ref.~\cite{lut02}. However, we note that in the
theoretical description of the $pp \to pp\eta$ reaction as reported in 
Refs~\cite{moa96,ged98}, the value of $a_R$ was similar to that used by us.
A smaller $a_R$ is also consistent with that extracted in Ref.~\cite{feu98}
in an effective Lagrangian model analysis of the meson-nucleon scattering.
Furthermore, it was noted in Ref.~\cite{gar02} that within a three-body 
model, the shapes of the $np \to \eta d $ cross sections can be explained
over a wide energy range only with a $a_R$ around 0.42 fm. A smaller value
of $a_R$ is also consistent with the J\"ulich model~\cite{gas03}. 
 
The FSI amplitude $T_{NN}$ has been calculated by following the Jost 
function method using the effective range expansion (ERE) of the $NN$ 
phase shifts as discussed in Refs.~\cite{wat69,shy98,shy99}. In case of 
the proton-proton sub-state, the Coulomb modified effective range expansion 
has been used~\cite{noy72}. The effective range parameters for the $NN$ 
channel have been taken to be the same as those used in Ref.~\cite{shy98}.
\begin{figure}
\begin{center}
\includegraphics[width=0.6 \textwidth]{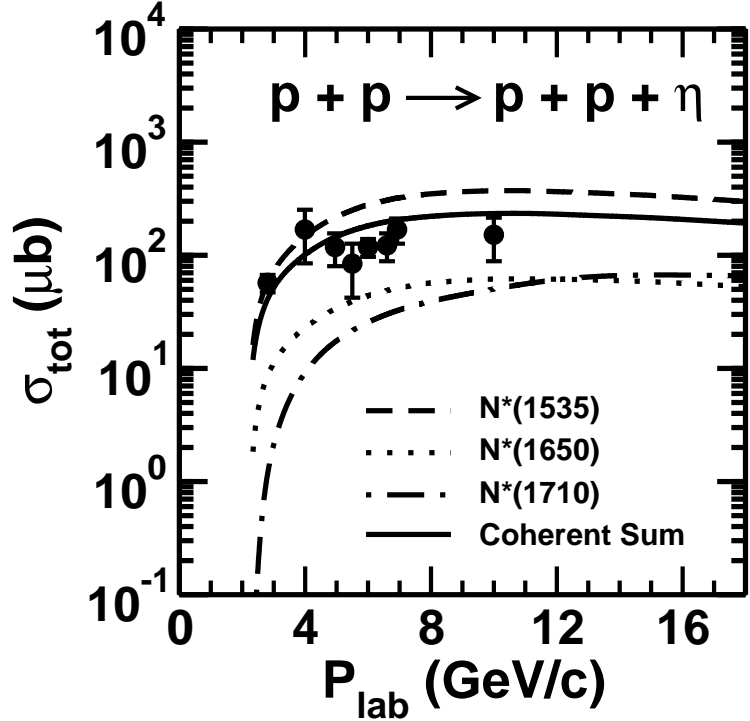}
\vskip -0.1in
\caption{
The total cross section for the
$p+p \rightarrow p+p+\eta$
reaction as a function of the beam momentum. The dashed, dotted and
dashed-dotted curves represent the contributions of $N^*$(1535),
$N^*$(1650), $N^*$(1710) baryonic resonance intermediate states, 
respectively. Their coherent sum is shown by the solid line. The 
experimental data are from~\protect\cite{lan88}. 
}
\end{center}
\end{figure}

It should however, be mentioned here that the use of on-shell forms to 
describe the FSI $T$ matrices $T_{if}$ has been criticized by some
authors. It has been argued in Refs.~\cite{han99,bar01} that the absolute
magnitudes of the cross sections obtained by such a procedure could
be uncertain because of the off shell effects. Even the Jost function 
method has been shown~\cite{gas05} to produce inadequate results in an 
study where the scattering length parameters for the $\Lambda - p$ final
state has been extracted from the $pp \to p\Lambda K^+$ data. In the 
next section we have examined the role of the off shell effects in the
$NN$ sub-state in some more details.
 
\section{results and discussions}

The major aim of this paper is to check the suitability of our model
and the vertex parameters appearing therein to describe the $\eta$ 
production cross sections over a wide range of beam energies. We have
therefore, applied our approach to describe the total cross sections for
the $pp \to pp\eta$ reaction for beam energies ranging from near threshold 
to upto 10 GeV, and for the $pn \to pn\eta$ reaction for beam energies 
from threshold to upto 1.6 GeV. These are the energy regimes in which 
experimental data are available for the two reactions. We have also used
this method to describe one set of the exclusive data, namely the 
$\eta$ angular distributions for the former reaction. Calculations have
been performed by using both the PS and PV couplings for $N^*N\pi$ and 
$N^*N\eta$ vertices. We note that the cross sections remain almost 
unchanged by switching from one type of coupling to another. In all the 
calculations shown below the coupling constants and cut off parameters 
for various vertices were the same as those discussed in section II. 

A cleaner check of the vertex parameters used in calculating the
amplitude $T_0(NN \rightarrow NN\eta)$, is provided by the comparison
of our calculations with the data for beam momenta above 3 GeV/c, since at
these energies FSI effects are most likely to be unimportant. In Fig.~3, we
show the comparison of our calculations and the experimental data (taken from
Ref.~\cite{lan88}) for the total cross section of the $pp \to pp\eta$ 
reaction at higher beam energies. We notice that the measured cross sections 
are reproduced reasonably well by our calculations (solid line) in the
entire range of beam momenta. 
\begin{figure}
\begin{center}
\includegraphics[width=0.6 \textwidth]{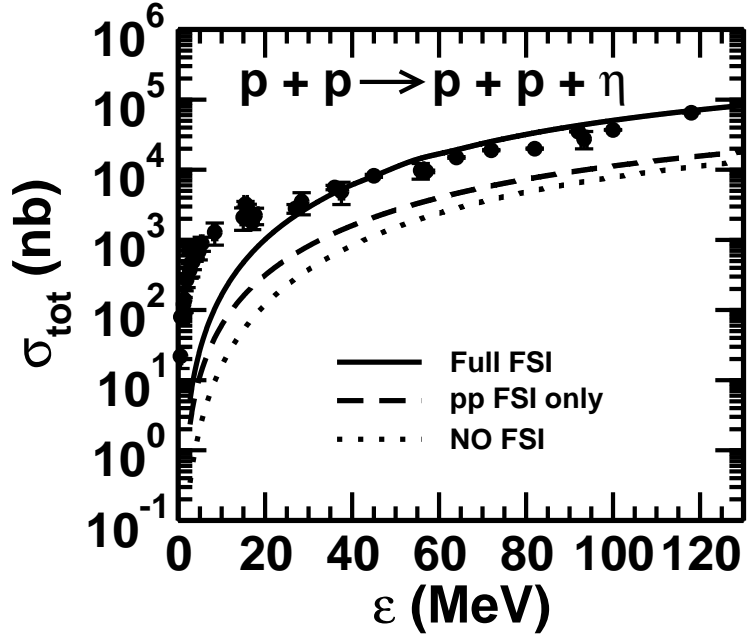}
\vskip -0.1in
\caption{
The total cross section for the $p + p \rightarrow p + p + \eta$
reaction as a function of the excess energy. The  dotted and
dashed curves represent cross section obtained with FSI effects included
only in the proton-proton sub-state of the final channel and no FSI at all,
respectively. The solid line shows the results obtained with full FSI effects
included as discussed in section II. The experimental data are 
from~\protect\cite{cal99,smy00,bar03,bal04,mos04}.
}
\end{center}
\end{figure}
 
Individual contributions of various nucleon resonance intermediate
states to the $pp\rightarrow pp\eta$ reaction are also shown in Fig.~3.
Cross sections corresponding to $N^*$(1535), $N^*$(1650) and $N^*$(1710)
resonances are represented by dashed, dotted and dashed-dotted lines, 
respectively while their coherent sum is shown by the solid line. We note that
the contributions of the $N^*$(1535) resonance dominate the total cross
section for all the beam momenta. In comparison, those of  $N^*$(1650) 
and $N^*$(1710) resonances are smaller by factors ranging from 5 - 10.  
However, the interference terms of the amplitudes corresponding to 
various resonances are not negligible. It must again be emphasized that 
we have no freedom of choosing the relative signs of the interference terms.

The results shown in Fig.~3 fix the parameters of all the vertices. In 
the application of our model to describe $NN\eta$ data at near 
threshold beam energies, the amplitude $T_0(NN \rightarrow NN\eta)$ has
been calculated with exactly the same values for all the parameters. For 
these energies the FSI effects in the outgoing channels, have been included
by using Eqs.~(27-30). The experimental cross sections in this energy regime 
are given as a function of the excess energy ($\epsilon$) which is defined
as $\epsilon = \sqrt{s} - 2m_p - m_\eta$, where $\sqrt{s}$ is the invariant
mass.
\begin{figure}
\begin{center}
\includegraphics[width=0.6 \textwidth]{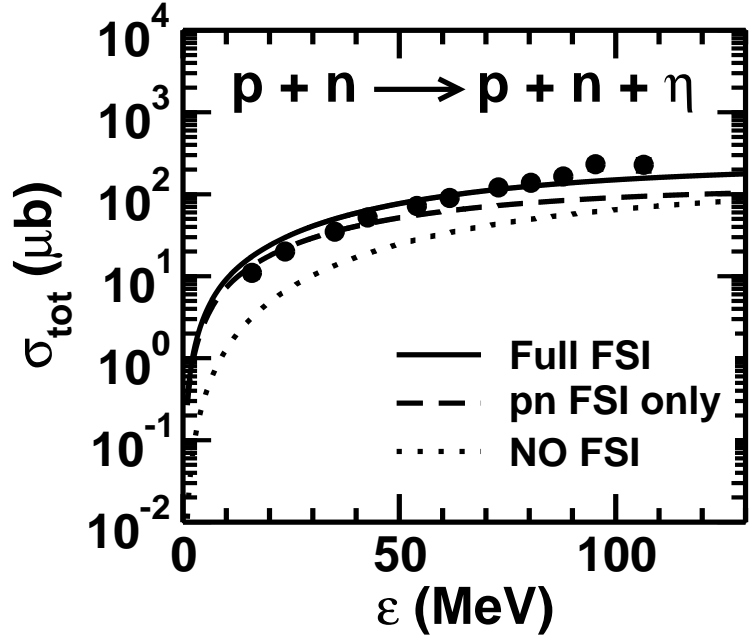}
\vskip -0.1in
\caption{
The total cross section for the $p + n \rightarrow p + n + \eta$
reaction as a function of the excess energy. The  dotted and
dashed curves represent cross section obtained with FSI effects included
only in the proton-proton sub-state of the final channel and no FSI at all,
respectively. The solid line shows the results obtained with full FSI effects
included as discussed in section II. The experimental data are 
from~\protect\cite{cal98}.
}
\end{center}
\end{figure}

In Figs.~4 and 5, we present comparisons of our calculations with the
experimental data for total cross sections of the $pp \to pp\eta$ and 
$pn \to pn \eta$ reactions, respectively, as a function of $\epsilon$. 
The solid, dashed and dotted lines show the results obtained by
including the full FSI effects in all the three sub-systems, FSI only in 
$pp$ and $pn$ channel and no FSI at all, respectively. It should be noted
that no arbitrary normalization constant has been introduced in any of the
results shown in these figures. For the case of the $pp \to pp\eta$ reaction,
our full calculations describe the data quite well for $\epsilon$ values in
the range of 15 - 130 MeV. On the other hand, for the  $pn \to pn \eta$
reaction they are in excellent agreement with the available data for all
the beam energies. The FSI in the $\eta p$ sub-state is indeed quite
important in our model. The difference between results obtained with FSI
in only the $pp$ sub-state and that in all the three subsystems of the final
channel, is comparable to that reported in the three body calculations of 
Ref.~\cite{fix04}. These authors have presented their results for 
$\epsilon$ values up to only 60 MeV. It would be interesting to see the
results of their model also at higher values of $\epsilon$. It should, 
however, be noted that the description of the data for the $pn \to pn \eta$
reaction within the three-body model model is less satisfactory in comparison
to that for the $pp \to pp \eta$ reaction. 

We see that the ELM is able to describe both the energy dependence and
the absolute magnitudes of the experimental cross sections for the $\eta$ 
meson production in both $pp$ and $pn$ channels for excess energies $>$ 15 
MeV. However, it underpredicts the $pp$ channel data for $\epsilon$ values
below 15 MeV. Such a trend has also been seen in calculations presented in 
Refs.~\cite{nak02,nak03} where the $\eta$ meson production in $NN$ 
collisions has been investigated within a relativistic meson exchange model
including the initial state interactions and FSI only in the $NN$ 
sub-system. These authors have attributed the near threshold underestimation
of the experimental total cross section to the non-inclusion of the 
$\eta p$ FSI in their model.
\begin{figure}
\begin{center}
\includegraphics[width=0.6 \textwidth]{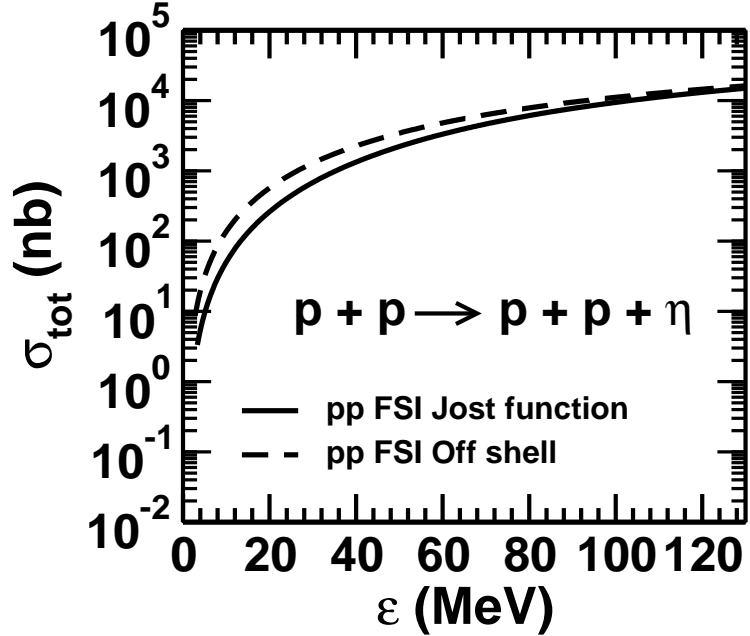}
\vskip -0.1in
\caption{
The total cross section for the $p + p \rightarrow p + p + \eta$
reaction as a function of the excess energy with only the $pp$ FSI
effects. The  solid and dashed curves represent cross section obtained 
with $pp$ FSI calculated with Jost function method and that with the method
described in Ref.~\protect\cite{del04}.  
}
\end{center}
\end{figure}

Since the ingredients of the primary production amplitude of our model have 
already been checked and fixed by calculations done at higher beam energies 
where FSI effects are absent, the underestimation of the $pp \to pp\eta$ cross
section for very low values of $\epsilon$ indicates that we need to improve
the treatment of the FSI effects. Inclusion of the off shell effects in 
the calculations of FSI is one of the likely improvements. The knowledge 
about the off shell nature of the $\eta N$ interaction is still very sparse.
However, we can use the results presented in Ref.~\cite{del04} to investigate
the effects of using off shell $pp$ FSI on the near threshold $\eta$ 
production cross sections. In Fig.~6, we have compared the results for the
total cross sections for the $pp \to pp\eta$ reaction obtained by including
FSI in $pp$ substate calculated within Jost function technique and that 
obtained with the method described in Ref.~\cite{del04} which includes
off-shell effects. In the results presented in this figure, no $\eta p$ 
FSI has been considered. We note that the off shell effects in the $pp$ 
FSI do increase the cross section for $\epsilon < 60$ MeV. However,
it is not enough to explain the underprediction of the experimental data by
our theory at smaller energies. One needs to have a better understanding of 
the $\eta N$ scattering amplitude. Further improvement may come by including
the three-body terms in the expansion of the scattering amplitude given by 
Eq.~(23). It has been shown in Ref.~\cite{fix04} that $NN\eta$ FSI effects 
calculated within a three-body scattering theory leads to enhanced 
cross sections at very low values of $\epsilon$. 
\begin{figure}
\begin{center}
\includegraphics[width=0.6 \textwidth]{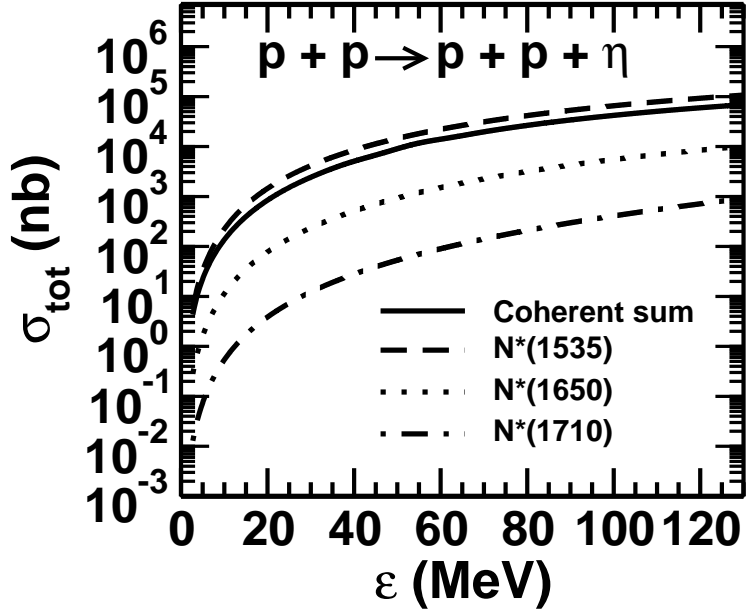}
\vskip -0.1in
\caption{
Contributions of $N^*(1535)$ (dashed line), $N^*(1650)$
(dotted) and $N^*(1710)$ (dashed-dotted line)
baryonic resonances to the total cross section for $pp \to pp \eta$
reaction. Their coherent sum is shown by the solid line.
}
\end{center}
\end{figure}

It should be remarked here that the differences in the cross sections of
$pn \to pn \eta$ and $pp \to pp \eta$ reactions are not only due to 
different isospin factors, but also due to differences in the FSI effects. 
The low energy scattering parameters between $pp$ and $pn$ cases are 
different; the latter involves also a triplet spin state together with the
singlet one. A crucial difference between them is the Coulomb interaction.
This is not included in the three-body model calculations of the 
$pp \to pp \eta$ reaction reported in Ref.~\cite{fix04}. Inclusion of this 
term is likely to reduce the cross section for beam energies very close to 
the threshold.

In Figs~7 and 8 we show the individual contributions of various nucleon
resonances to the total cross sections of the $pp \to pp\eta$ and 
$pn \to pn\eta$ reactions, respectively, at the near threshold beam energies.
Similar to the situation at higher beam energies, the cross sections are 
dominated by the $N^*$(1535) resonance excitation. Since $N^*$(1535) is the
lowest energy baryonic resonance having an appreciable branching ratio for 
the decay into the $N\eta$ channel, its dominance in this reaction even at 
beam energies near the $\eta$ production threshold is to be expected.
The contribution of $N^*$(1650) resonance state is small and that of the  
$N^*$(1710) resonance is even smaller at these lower beam energies. It should,
however be noted that the resonance-resonance interference terms are not 
negligible. For the $pp \to pp\eta$ reaction, the total cross sections are  
smaller than those corresponding to  the $N^*$(1535) resonance alone.
For the $pn \to pn \eta$ reactions the difference between the two is not
visible in Fig.~8. 
\begin{figure}
\begin{center}
\includegraphics[width=0.6 \textwidth]{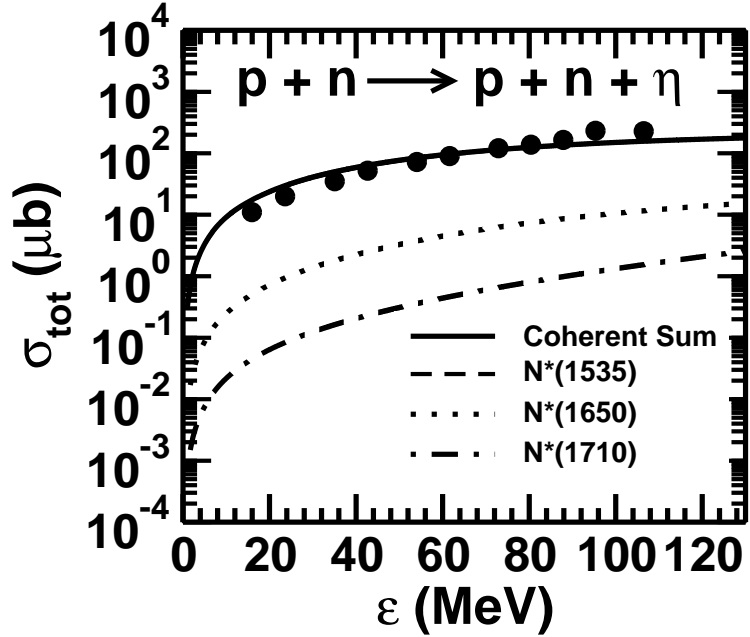}
\vskip -0.1in
\caption{
Contributions of $N^*(1535)$ (dashed line), $N^*(1650)$
(dotted) and $N^*(1710)$ (dashed/dotted line)
baryonic resonances to the total cross section for $pn \to pn \eta$
reaction. Their coherent sum is shown by the solid line.
}
\end{center}
\end{figure}
 
We found that the inclusion of the amplitudes corresponding to the 
nucleon intermediate states (the nucleon bremsstrahlung) made a negligible
difference in the results reported in Figs.~1-8 if the value of the 
coupling constant $g_{NN\eta}$ is taken below 3.0. With the largest
value of $g_{NN\eta}$ used in the literature (6.14), the 
results were affected to the extent of only a few percent. This result is in
agreement with that reported in Ref.~\cite{nak02}. It is obvious that due
to a considerable amount of uncertainty in the value of $g_{NN\eta}$ (see, 
e.g.~\cite{ben95,ose94,tia02}), the nucleon excitation amplitudes are quite 
uncertain (see, e.g.~\cite{ben95,ose94,tia02}) and their inclusion makes an 
insignificant difference to the results reported above. 

In Fig.~9, we show contributions of various meson exchanges to the
$pp \to pp \eta$ reaction at near threshold beam energies. The dashed,
dotted, dashed-dotted and dashed-double dotted curves represent the 
contributions of $\pi$, $\rho$, $\sigma$ and $\omega$ meson exchanges, 
respectively. Their coherent sum is shown by the solid line. We see that
the one pion exchange graphs make the largest contribution to the reaction
in this energy regime. However, a striking feature of this figure is
that despite a larger value for the $g_{N^*N\rho}$ coupling used in our 
calculations, the contributions of $\rho$ meson exchange is still much smaller
than that of the pion exchange graphs. Hence in contrast to the results 
reported in Refs.~\cite{ged98,san98} the $\rho$ meson exchange terms do not
dominate the total $NN\eta$ production cross sections. To understand this 
difference we note that while in Refs.~\cite{ged98,san98} $\gamma_5 \gamma_\mu$
couplings have been used for the $\rho NN^*$ vertex, we have taken a 
$\gamma_5\sigma_{\mu \nu}$ coupling for the same which is an extension of the
$\gamma NN^*$ couplings (due to vector meson dominance reasons~\cite{ben95}). 
This is also compatible with the forms of the $\rho NN^*$ couplings used in 
the literature~\cite{pos01,ris01,pen02}. Since the $\rho$ meson exchange 
amplitudes calculated with the $\gamma_5\sigma_{\mu \nu}$ couplings involve
delicate cancellations among leading terms, contributions of this exchange 
diagrams to the $\eta$ production cross sections are weakened. This is the 
main reason for differences between our results and those of 
Refs.~\cite{ged98,san98}. 

In Ref~\cite{nak02}, although the form of the
$\rho NN^*$ coupling is the same as ours, relatively lower $\rho$ meson
exchange cross sections result from the use of a very small value for the
coupling constant $g_{N^*N\rho}$ which is based on the lower limit of the 
branching ratio for the radiative decay of this resonance. Our value for
this constant, on the other hand,  is calculated from the branching ratio of
the decay of this resonance to $N\rho$ channel as quoted in Ref.~\cite{pdg06}.
\begin{figure}
\begin{center}
\includegraphics[width=0.5 \textwidth]{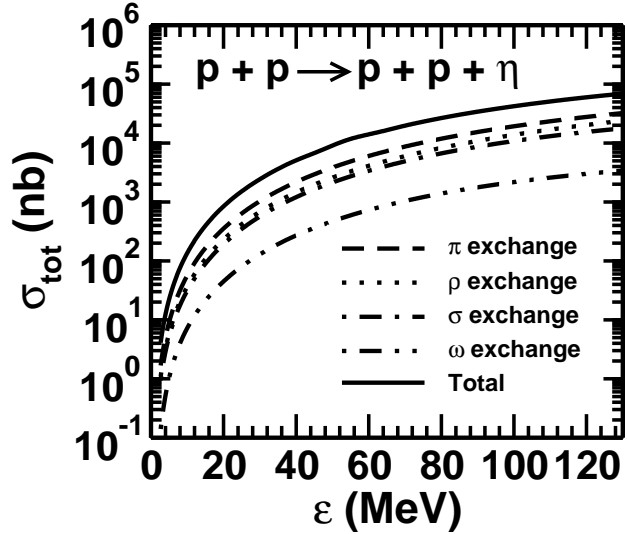}
\vskip -0.1in
\caption{
Contributions of $\pi$ (dashed line), $\rho$ (dotted line), $\omega$ 
(dashed-double dotted line) and $\sigma$ (dashed-dotted line)
meson exchange processes to the total cross section for the 
$pp \to pp \eta$ reaction as a function of the excess energy.
Their coherent sum is shown by the solid line. 
}
\end{center}
\end{figure}
 
We see that the $\omega$ meson exchange process contributes insignificantly
to the $NN\eta$ production but the $\sigma$ meson exchange terms are 
relatively more important. Larger contributions from the latter has also
been seen in other subthreshold reactions analyzed within our model. It
indicates that $\sigma$ meson exchange may be an efficient means of 
mediating the large momentum mismatch involve in the meson production in
$NN$ collisions~\cite{lee93,hor94}.  
    
In fig.~10, we investigate the effects of using PS or PV couplings for
the $N^*N\pi (\eta)$ vertices. We notice hardly any difference in the 
cross sections calculated by the two types of couplings. Similar results
were also observed in Ref~\cite{pen02}. This result is not surprising 
since the two couplings are constructed in such a way that both are
equivalent on the mass shell. Of course, they start having different
energy behavior in the far off shell region where resonance contribution
is anyway suppressed due to dominance of the corresponding propagator.
It is only in the $NN\pi$ case that difference in the PS and PV couplings
are obvious with a clear preference for the PV coupling in line with the
chiral symmetry as discussed earlier.
\begin{figure}
\begin{center}
\includegraphics[width=0.6 \textwidth]{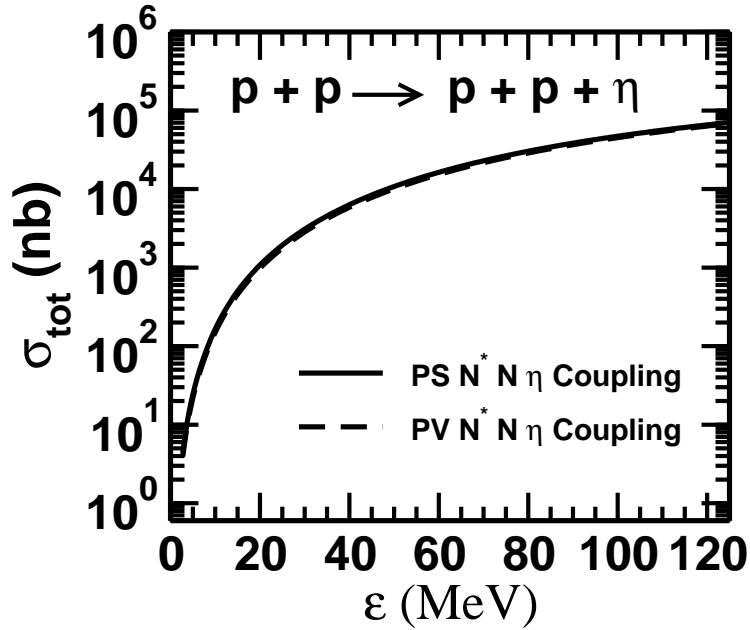}
\vskip -0.1in
\caption{
The total cross section for the $pp \to pp \eta$ reaction calculated with
pseudovector (solid line) and psuedoscalar (dashed line) couplings for the 
$N^*N\pi(\eta)$ vertex for the resonances considered in this paper, as a 
function of the excess energy.
}
\end{center}
\end{figure}

After establishing the dynamical content of our model {\it vis a vis} the 
description of the total production cross sections, we now turn our attention
to more exclusive data. In Fig.~11, we show a comparison of our calculations 
with the data for the angular distribution of $\eta$ meson in the 
$pp \to pp \eta$ reaction for $\epsilon$ values of 15 MeV (upper panel)
and 41 MeV (lower panel). Since the angular distribution data 
are normalized to the total cross sections for both values of $\epsilon$,
we have done the same in our calculations shown in this figure. We note that
shapes of the angular distributions are described well by our model 
at both the energies. At the lower beam energy, the data as well as our 
calculations have essentially isotropic distributions. However, for 
$\epsilon$ = 41 MeV, there is a tendency in our calculations to show slight 
enhancements at forward and backward angles which is typical of the $\pi$ 
exchange dominance process in the $N^*(1535)$ excitation. Due to
large statistical errors in the data, it is difficult to conclude if they  
show a trend different from our calculations. It will be useful to have  
better quality data with less statistical errors in order to determine
if other mechanisms which may show a trend different from ours, are also 
important. In any case, it is quite unlikely that the $\rho$ meson exchange
mechanism which might lead to a distribution different from that seen in our 
calculations~\cite{fal01} is a dominant mechanism as has already been 
discussed.
\begin{figure}
\begin{center}
\includegraphics[width=0.5 \textwidth]{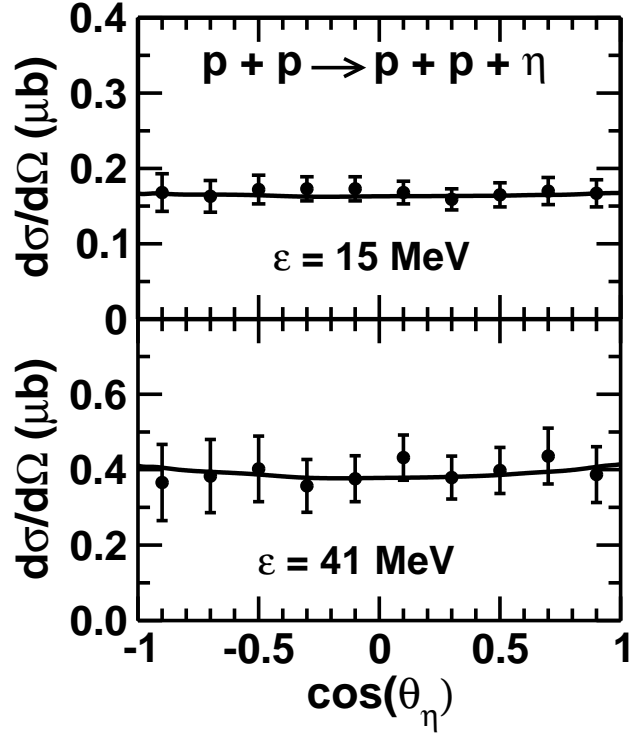}
\vskip -0.1in
\caption{
Differential cross sections of the $pp \to pp \eta$ reaction as a 
function of $\eta$ meson angle in the c.m. frame of the total system 
at the excess energies of 15 MeV (upper panel) and 41 MeV (lower panel).
The experimental data have been taken from the Ref.~\protect\cite{bar04}.
}
\end{center}
\end{figure}

\section{Summary and Conclusions}
In this paper we investigated the $\eta$ meson production in proton-proton
and proton-neutron collisions for beam energies ranging from near threshold
to about 10 GeV within an effective Lagrangian model which has been used  
previously to describe successfully the pion, associated kaon and dilepton 
production in $NN$ collisions. The interaction between two nucleons in the
initial state is modeled by the effective Lagrangians based on the exchange
of $\pi$, $\rho$, $\omega$ and $\sigma$ mesons. The parameters of the 
corresponding vertices were taken to be the same as those used in the 
previous applications of this model which restricts the freedom of varying
the parameters to get fits to the data. The eta meson production proceeds
via excitation, propagation and decay of $N^*(1535)$, $N^*(1650)$ and 
$N^*(1710)$ intermediate nucleon resonance states. The coupling constants 
at the resonances-nucleon-meson vertices have been determined from the 
experimental branching ratios of the decays of the resonances into the relevant
channels. Here again we have used the same coupling constants as those used
in the previous applications of the model at vertices that appeared also in
those calculations. The interference terms among various amplitudes are 
included in the total transition matrix.
 
To describe the data at the near threshold beam energies, the FSI effects
among the outgoing particles are included by a generalized Watson-Migdal
method which allows to have these effects in all the three two body
sub-systems of the out-going channel. This method involves a parameter
which has been determined from the constrained that in the limit of FSI 
in only $NN$ sub-system the result of the usual Watson method is reproduced.

In this paper we presented the analysis of the data for total cross sections
for the $pp \to pp\eta$ and $pn \to pn\eta$ reactions and for the $\eta$ 
angular distributions in the former reaction. With the same set of vertex 
parameters, the model is able to provide a good description of the data for
the $pp \to pp \eta$ reaction at higher as well as near threshold beam 
energies except for the excess energies below 15 MeV where our calculations
underpredict the experimental data. The experimental total cross sections of
the $pn \to pn \eta$ reaction are also well described by our model. The data
for the $\eta$ angular distributions in the case of $pp \to pp \eta$ reaction
are also well reproduced at two beam energies. Imprecise knowledge of the
$\eta N$ scattering amplitudes and non-inclusion of the three-body effects 
are the most likely reasons for underestimation of the $pp$ channel data by
our model at very low beam energies.

In this study we have not studied the observables related to meson energy 
dependence e.g., final $pp$ and $\eta p$ invariant mass distributions. 
There are some open theoretical issues concerning the explanation of the 
corresponding experimental data. While in Ref.~\cite{nak03}, the inclusion
of contributions of the non-$s$-wave states in the $pp$ subsystem were 
found to be essential to explain these data, the three-body effects in the 
$pp\eta$ system and not the contribution of the higher partial waves 
were shown to be crucial for this purpose by the authors of 
Ref.~\cite{fix04}. At this stage, our theory excludes both these effects.
Extension of our model to include these mechanisms is vital before we can
make some meaningful contribution towards settlement of this issue.  

Within our model, one pion exchange processes make the largest 
contributions to cross sections in the entire energy regime. Despite our 
using a large coupling constant for the $N^*(1535)N\rho$ vertex, the 
cross sections of the $\rho$ meson exchange process are still lower than
those of the pion exchange mechanism. Therefore, $\rho$ meson exchange being
the dominant mechanism of $N^*(1535)$ resonance excitation~\cite{fal01} is 
not supported by our calculations. The individual contributions of the 
$\omega$ meson exchange diagrams are very small every where. On the other
hand, the $\sigma$ exchange terms make relatively larger contributions.   

The excitation of the $N^*(1535)$ resonance dominates the $NN\eta$ 
production at both higher as well as  near threshold beam energies.
The contributions of $N^*(1650)$ and $N^*(1710)$ are small in comparison.
However, the interference among various resonance contributions is not
negligible. Unlike the $NN\pi$ vertex where there is a clear preference
for the PV coupling, the present reaction does not distinguish between PS and
PV couplings at the $N^*N\eta$ vertex involving spin-1/2 even or odd parity
resonances. We point out that the mechanism of the $pp\eta$ production via 
preferential excitation of the $N^*(1535)$ intermediate baryonic resonance
state in one-pion-exchange process has received support recently from an 
experimental study~\cite{czy07} of the analyzing powers of the 
${\bar p} + p \to p + p + \eta$ reaction.   

This work fixes the parameters of the effective Lagrangian model for most of the
vertices involve in the eta meson production processes. An interesting further 
check of this model will be provided by the analysis of the eta photoproduction 
data on nucleons (see,e.g.~\cite{kuz06,kru06}). An exciting recent result is 
that the integrated cross section of the photoproduction of $\eta$ 
meson on neutrons shows an additional maximum at center of mass energies 
around 1.66 GeV. This has recently been explained in terms of the excitation
of the $N^*(1650)$ and $N^*(1650)$ resonance states~\cite{shk06}. 
Furthermore, the vertex parameters derived by us will also be useful in 
applications of effective Lagrangian method in describing the production
of eta-mesic nuclei in proton and photon induced reactions 
(see, e.g.~\cite{ani04,bas03}).

\section{Acknowledgments}
The author is thankful to Andrzej Deloff for several useful correspondences
regarding the off shell $pp$ FSI method presented in Ref.~\cite{del04} and for
his help in implementing the calculations based on this method. Useful
discussions with Horst Lenske, Jerry Miller, Ulrich Mosel, Pawel Moskal and 
Madeleine Soyeur is gratefully acknowledged. 
\appendix
\section{Final state interaction amplitudes for three-particle states}

We give here some clarifications and steps leading to the derivation Eq.~(23).

The total Hamiltonian of the three-particle system is written as
$H  =  H_0 + U$ where $H_0$ is the kinetic energy operator of the system
and the interaction $U$ is taken as $U  =  U_{23} + U_{31}
+ U_{12} \equiv U_1 + U_2 + U_3$ assuming that the three-particle
states interact by means of the additive pair interactions represented by 
$U_k$. The Green's functions corresponding to $H$ and $H_0$ 
are, respectively
\begin{eqnarray}
G^{(\pm)} (E) & = & lim_{\epsilon \rightarrow 0}\, {
           {1}\over {E - H_0 - U \pm i\epsilon}},\\
G_0^{(\pm)} (E) & = & lim_{\epsilon \rightarrow 0}\, {
           {1} \over {E - H_0 \pm i\epsilon}}
\end{eqnarray}
We shall also need the Hamiltonian describing the two particles interacting
while the third one is free, namely, $H_k = H_0 + U_k$ and the corresponding
Green's functions 
\begin{eqnarray}
G_k^{(\pm)} (E) & = & lim_{\epsilon \rightarrow 0}\, {
           {1}\over {E - H_0 - U_k \pm i\epsilon}}
\end{eqnarray}
The full three-body transition operator $T$ satisfies the Lipmann-Schwinger 
equations 
\begin{eqnarray}
T(E) & = & U + U G_0 T(E), 
\end{eqnarray}
which can also be written as
\begin{eqnarray}
T(E) & = & \sum_k\,U_k + \sum_k\,U_k G_0 T(E), 
\end{eqnarray}
Eq.(A5) leads to the iteration
\begin{eqnarray}
T(E) & = & \sum_k\, [U_k + U_k G_0 U_k + U_k G_0 U_k G_0 U_k +...] \nonumber \\
     && + \sum_{k \not= j}\, [U_k + U_k G_0 U_k + ...]\, G_0\,
        [U_j + U_j G_0 U_j + ...] + ...
\end{eqnarray}
 
In the study of the final state interaction problems, one usually has 
in addition to potential U [which is responsible for the transition from
the initial state to the free final states characterized by the transition 
matrix $T_0$ in Eq. (22)], an additional interaction V that describes a 
type of internal interaction among the constituents of the final state. 
The total Hamiltonian is then written as $H_0$ + $U$ + $V$ and one can make
use of the standard "two-potential formalism" of Goldberger and Watson
(see, e.g., \cite{wat69}) to write the total scattering matrix element 
$(T_{fi})$ as as a sum of two terms - one of them involve the matrix
elements of the interaction $U$ between the exact initial state wave function 
and the final scattering state wave function corresponding to interaction
$V$. From the general theory, we can write 
\begin{eqnarray}
T_{fi} & = & < \chi_f |U + V| \psi_i^{(+)} >,
\end{eqnarray}
where $\psi_i^{(+)}$ completely describes the initial state with the
outgoing wave boundary condition, and $\chi_f$ is the final plane wave 
state. One can eliminate $\chi_f$ by introducing wave functions 
$\phi_f^{(-)}$ which are the eigenfunctions of the Hamiltonian $H_0 + V$ as  
\begin{eqnarray}
\phi_f^{(-)} & = & \chi_f + G_0^{(-)} V \phi_f^{(-)}
\end{eqnarray}
Substituting (A8) into (A7) one gets after some manipulation,
\begin{eqnarray} 
T_{fi} = < \phi_f^{(-)} |U| \psi_i^{(+)} > + < \phi_f^{(-)} |V| \chi_i >,
\end{eqnarray}
where $\chi_i$ is the initial plane wave state. The final state interactions
of interest are contained in $\phi_f^{(-)}$.

In applications of relevance to us, the second term of Eq.~(A9) would
vanish because we assume $V$ can not create real mesons. Eq.~(23) can be
obtained by using an iteration similar to that given by 
Eq.~(A6) in the remaining (first) term of Eq.~(A9). The amplitude $T_0$
is defined in the same way as the first term of Eq.~(A9) with a plane wave
final state. The amplitude $T_k$ in the second term of Eq.~(23) represents
the matrix elements of the interaction $V_k$ between the plane wave and the
scattering states of particle $ij$. We define $V_k$ for the partition $k$ in
the same way as $U_k$.

\end{document}